\documentclass{achemso}
\usepackage[T1]{fontenc} 
\usepackage{graphicx} 
\usepackage{amsmath}  
\usepackage{indentfirst} 
\usepackage{float}    
\usepackage{etoolbox} 
\usepackage{fancyhdr} 
\usepackage[colorlinks=true,linkcolor=blue,citecolor=blue,urlcolor=blue]{hyperref} 
\usepackage{fancyhdr} 
\usepackage{lastpage} 
\clearpage
\usepackage{xcolor}
\usepackage{arabtex}  
\usepackage{sectsty}
\usepackage{etoolbox}
\usepackage[final]{changes}

\usepackage[pagewise]{lineno}

\newcommand{\colortext}[1]{{\color{black}#1}} 

\newcommand{\supplementaryinfo}{\href{http://pubs.acs.org}{Supporting Information}}
\setkeys{acs}{etalmode = truncate, maxauthors=3} 
\makeatletter
\renewcommand*{\acs@author@fnsymbol@symbol}[1]{
    \ifcase #1 *\or
    1\or
    2\or
    3\or
    4\or
    5\or
    6\or
    7\or
    8\or
    9\or
    10
    \fi
}
       
\renewcommand*\acs@contact@details{
    {\sffamily *\,E-mail: \acs@email@list }%
    \acs@number@list
}           

\patchcmd{\acs@address@list@auxii}
{\acs@author@fnsymbol{\acs@affil@marker@cnt}}
{\textsuperscript{\acs@author@fnsymbol{\acs@affil@marker@cnt}}}
{}{}

\patchcmd{\acs@address@list@auxii}
{{\acs@author@fnsymbol{\acs@affil@marker@cnt}\@nameuse{@altaffil@\@roman\@tempcnta}\par}}
{{\textsuperscript{\acs@author@fnsymbol{\acs@affil@marker@cnt}}\@nameuse{@altaffil@\@roman\@tempcnta}\par}}
{}{}        

\makeatother
\author{Ning Wang}
\affiliation[Key Laboratory]
{CAS Key Laboratory of Quantum Information, University of Science and Technology of China, Hefei, Anhui 230026, China}
\alsoaffiliation[CAS Center]
{CAS Center for Excellence in Quantum Information and Quantum Physics, University of Science and Technology of China, Hefei, Anhui 230026, China}

\author{Jia-Min Kang}
\affiliation[Key Laboratory]
{CAS Key Laboratory of Quantum Information, University of Science and Technology of China, Hefei, Anhui 230026, China}
\alsoaffiliation[CAS Center]
{CAS Center for Excellence in Quantum Information and Quantum Physics, University of Science and Technology of China, Hefei, Anhui 230026, China}

\author{Wen-Long Lu}
\affiliation[Key Laboratory]
{CAS Key Laboratory of Quantum Information, University of Science and Technology of China, Hefei, Anhui 230026, China}
\alsoaffiliation[CAS Center]
{CAS Center for Excellence in Quantum Information and Quantum Physics, University of Science and Technology of China, Hefei, Anhui 230026, China}

\author{Shao-Min Wang}
\affiliation[Key Laboratory]
{CAS Key Laboratory of Quantum Information, University of Science and Technology of China, Hefei, Anhui 230026, China}
\alsoaffiliation[CAS Center]
{CAS Center for Excellence in Quantum Information and Quantum Physics, University of Science and Technology of China, Hefei, Anhui 230026, China}

\author{You-Jia Wang}
\affiliation[Key Laboratory]
{CAS Key Laboratory of Quantum Information, University of Science and Technology of China, Hefei, Anhui 230026, China}
\alsoaffiliation[CAS Center]
{CAS Center for Excellence in Quantum Information and Quantum Physics, University of Science and Technology of China, Hefei, Anhui 230026, China}

\author{Hai-Ou Li}
\affiliation[Key Laboratory]
{CAS Key Laboratory of Quantum Information, University of Science and Technology of China, Hefei, Anhui 230026, China}
\alsoaffiliation[CAS Center]
{CAS Center for Excellence in Quantum Information and Quantum Physics, University of Science and Technology of China, Hefei, Anhui 230026, China}
\alsoaffiliation[National Laboratory]
{Hefei National Laboratory, University of Science and Technology of China, Hefei 230088, China} 

\author{Gang Cao}
\affiliation[Key Laboratory]
{CAS Key Laboratory of Quantum Information, University of Science and Technology of China, Hefei, Anhui 230026, China}
\alsoaffiliation[CAS Center]
{CAS Center for Excellence in Quantum Information and Quantum Physics, University of Science and Technology of China, Hefei, Anhui 230026, China}
\alsoaffiliation[National Laboratory]
{Hefei National Laboratory, University of Science and Technology of China, Hefei 230088, China}

\author{Bao-Chuan Wang}
\affiliation[Key Laboratory]
{CAS Key Laboratory of Quantum Information, University of Science and Technology of China, Hefei, Anhui 230026, China}
\alsoaffiliation[CAS Center]
{CAS Center for Excellence in Quantum Information and Quantum Physics, University of Science and Technology of China, Hefei, Anhui 230026, China}
\email{bchwang@ustc.edu.cn}

\author{Guo-Ping Guo}
\affiliation[Key Laboratory]
{CAS Key Laboratory of Quantum Information, University of Science and Technology of China, Hefei, Anhui 230026, China}
\alsoaffiliation[CAS Center]
{CAS Center for Excellence in Quantum Information and Quantum Physics, University of Science and Technology of China, Hefei, Anhui 230026, China}
\alsoaffiliation[National Laboratory]
{Hefei National Laboratory, University of Science and Technology of China, Hefei 230088, China}
\alsoaffiliation[Origin]
{Origin Quantum Computing Company Limited, Hefei, Anhui 230026, China}
\title
  {Highly tunable 2D silicon quantum dot array with coupling beyond nearest neighbors}
\abbreviations{QD,SET,DQD,TQD,QQD,2D,SEM}

\begin{document}


\begin{abstract}
  Scaling up quantum dots to two-dimensional (2D) arrays is a crucial step for advancing semiconductor quantum computation. However, maintaining excellent tunability of quantum dot parameters, including both nearest-neighbor and next-nearest-neighbor couplings, during 2D scaling is challenging, particularly for silicon quantum dots due to their relatively small size. Here, we present a highly controllable and interconnected 2D quantum dot array in planar silicon, demonstrating independent control over electron fillings and the tunnel couplings of nearest-neighbor dots. More importantly, we also demonstrate the wide tuning of tunnel couplings between next-nearest-neighbor dots, which plays a crucial role in 2D quantum dot arrays. This excellent tunability enables us to alter the coupling configuration of the array as needed. These results open up the possibility of utilizing silicon quantum dot arrays as versatile platforms for quantum computing and quantum simulation.

\end{abstract}

Semiconductor quantum dots, with their small footprints\cite{RN5361,RN11714,RN7360}, compatibility with semiconductor manufacturing processes\cite{RN10755,RN8628,RN10696,RN11981}, and the potential for scaling up\cite{RN1301,RN8678,RN1428}, have emerged as prime candidates for quantum computing and simulation. Recent achievements towards fault-tolerant quantum computing include single- and two-qubit gate fidelities exceeding the threshold\cite{RN10511,RN10510,RN10509,RN10653} required by the surface code\cite{RN10693,RN11492}, universal control of up to six qubits in linear arrays\cite{RN9013,RN11423,RN11493,RN11727}. Additionally, quantum dot systems, offering extensive and electrical in-situ tunability of device parameters, have demonstrated their capabilities for quantum simulation, including studies of Mott-Hubbard physics\cite{RN11032,RN1437,RN11790} and quantum magnetism\cite{RN10477} in linear QD arrays. To fully leverage these advantages, scaling up quantum dots to 2D arrays in a controllable manner is imperative. However, the small size of quantum dots poses significant constraints on their design, fabrication, and tunability. A critical obstacle in scaling up is maintaining excellent control over parameters such as electron filling, nearest-neighbor coupling, and particularly next-nearest-neighbor coupling, which is crucial in 2D quantum dot arrays. For instance, the next-nearest-neighbor coupling should be completely suppressed for the surface code error correction scheme relying on nearest-neighbor coupling\cite{RN10693,RN11492}; on the other hand, it can be introduced for simulations of exotic phases of matter such as quantum spin liquids\cite{RN11994,RN11972,RN11973} and high-temperature superconductivity\cite{RN11851,RN11866,RN11868}.

In recent years, 2D quantum dot arrays have been demonstrated in various semiconductor materials, including GaAs, planar Ge, and $^{31}\mathrm{P}$ donors. These arrays have shown significant potential for advancing quantum computing\cite{RN8629,RN8628,RN8962,RN11758,RN11982} and serving as small-scale simulators for exploring condensed matter physics\cite{RN7349,RN11787,RN11524,RN10935}. In the case of silicon quantum dots, challenges in device fabrication—particularly due to stronger electrostatic confinement—have led to few reports on 2D arrays\cite{RN9233,RN8628,RN11799}, which primarily focus on the tunability of tunnel coupling between nearest neighbors. While a CB gate has been proposed in GaAs QD array to suppress next-nearest neighbor couplings, its control capabilities have not been thoroughly studied, especially in planar silicon. Therefore, achieving a well-defined 2D quantum dot array with coupling beyond nearest neighbors requires not only careful gate architecture design but also comprehensive characterization methods.

In this article, we present a highly controllable and interconnected \(2 \times 2\) silicon quantum dot array \added{incorporating next-nearest-neighbor couplings with a wide tunable range,}\deleted{with the capability for selective control over tunnel couplings beyond nearest neighbors,} positioning it as a versatile platform for quantum applications. The device, fabricated on an undoped Si/SiGe heterostructure, comprises four quantum dots arranged in a square lattice and two larger quantum dots serving as charge sensors. We first demonstrate that all dots can be depleted to the single-electron occupancy. To intuitively acquire the electronic occupation and coupling state of all dots, we developed a synchronized sweeping method, achieving homogeneous and deterministic electron filling across the array. Next, we characterized the tunnel coupling of nearest-neighbor and next-nearest-neighbor dots, which can be independently tuned over a wide range. Finally, leveraging the device's remarkable tunability, we strategically modified the system's coupling configuration by deactivating certain couplings and discussed the potential applications for quantum computing and simulation.

\begin{figure}[H]
  \centering
  \includegraphics{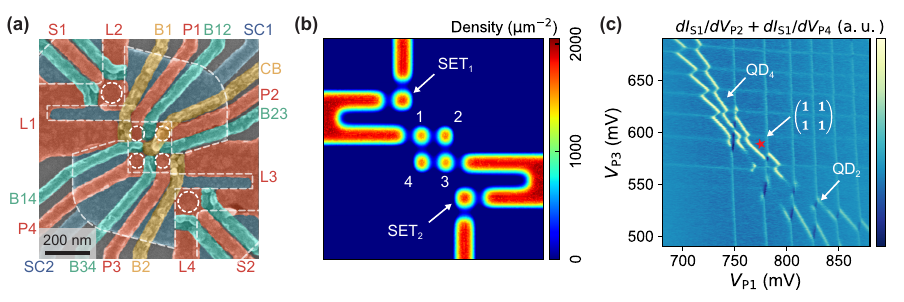}
  \caption{(a) False-colored SEM image of a device identical to the one used in the experiments. The device consists of four quantum dots formed beneath plunger gates P1-P4 (orange) and two proximal quantum dots formed beneath gates S1 and S2 as charge sensors. The white dotted circles indicate the formation locations of quantum dots, consistent with (b) the electron density in the quantum well simulated using the self-consistent Schrödinger-Poisson method. Four quantum dots are arranged in a square grid, labeled 1 to 4 clockwise. Dots 1 and 3 are directly coupled to the reservoirs, with barrier gates B1 and B2 (yellow) controlling the tunnel rates. Barrier gates $\mathrm{B}_{ij}$  (green) control the tunnel couplings between nearest-neighbor dots while the CB gate (yellow) controls the tunnel coupling between next-nearest-neighbor dots. (c) Charge-stability diagram for the array. Four sets of charge transitions with different slopes are visible, each corresponding to a different quantum dot. The array can be verified to reach the single-electron occupation state in each dot, marked by a red star.}
  \label{fig1:Figure}
\end{figure}

\autoref{fig1:Figure}a shows a false-colored scanning electron microscope (SEM) micrograph of a lithographically identical device used in this experiment. The corresponding electron density at the top plane of the quantum well is shown in \autoref{fig1:Figure}b. Electrostatic confinement is achieved by a multilayer aluminum gate stack\cite{RN1350,RN11424,RN10835} fabricated on an undoped Si/SiGe heterostructure. A \(2 \times 2\) QD array is formed underneath the plunger gates (P1-P4) with barrier gates (B12, B23, B34, B41) controlling the tunnel couplings between nearest-neighbor dots. Dots 1 and 3 are directly coupled to the reservoirs, while dots 2 and 4 are (un)loaded via the co-tunneling process. A central barrier gate (CB) extends toward the center of the array to control the tunnel couplings between dots 1(2) and 3(4), which has been mentioned in previous semiconductor QD arrays\cite{RN11799,RN1426}. Furthermore, two larger quantum dots are strategically positioned at the upper-left and lower-right corners of the array to serve as single-electron transistor (SET) charge sensors. Notably, the two SETs and the QD array share part of the reservoirs (L1 and L3), thereby reducing the complexity of device design and fabrication (see \supplementaryinfo~Section S1 for more details).

The device is characterized in a dilution refrigerator with a base temperature of about 20 mK. From the linewidth of the Coulomb blockade peak at zero bias, the QD electron temperature is estimated to be approximately 150 mK. Charge sensing is achieved by modulating the plunger gates above the quantum dots and measuring the transconductance of two charge sensors simultaneously. Unless otherwise specified, the AC excitations are typically applied to plunger gates P2 or/and P4 at a frequency of 73 Hz, while the integration time for current acquisition is set to 300 ms. To initially tune the device, we employed a consecutive tuning procedure adopted from ref.\citenum{RN1414}, transitioning from a double quantum dot (DQD) to a well-defined triangular triple quantum dot (TQD), and then to a square quadruple quantum dot (QQD) array (see \supplementaryinfo~Section S2). These preliminary results demonstrated the high degree of tunability and versatility in the gate architecture design. 

\autoref{fig1:Figure}c shows the charge-stability diagram for the array plotted as a function of $V_\mathrm{P1}$ and $V_\mathrm{P3}$, displaying four sets of charge transitions with different slopes. The horizontal and vertical transitions correspond to diagonal dots (dots 1 and 3), while the two inclined ones belong to anti-diagonal dots (dots 2 and 4). By counting the charge transitions from the lower-left corner in the diagrams, we can achieve deterministic filling of electrons in each dot and reach the $(N_1, N_2, N_3, N_4) = (1, 1, 1, 1)$ charge state (red star in \autoref{fig1:Figure}c), where $N_i$ indicates the electron number in dot $i$. In addition, the orthogonal charge transitions and right-angle intersections for diagonal dots suggest considerably weak inter-site Coulomb interaction and tunnel coupling. The symmetric coupling of dot 2(4) with dots 1 and 3, as evidenced by its charge transition with slope near -1, also implies significant capacitive crosstalk, potentially hindering precise parameter tuning.

\begin{figure}[H]
  \centering
  \includegraphics{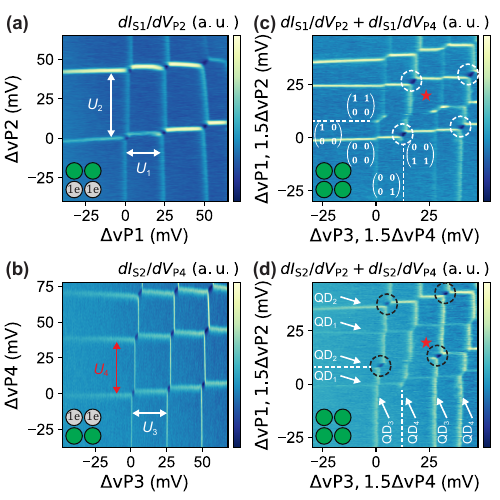}
  \caption{Homogeneous electron filling of the array. (a) and (b) Charge-stability diagrams for dot pairs 1-2 and  3-4, measured by sweeping virtual plunger gates. The colored circles in the lower-left corner of the diagrams represent the occupation state of the array. Green indicates the dot being swept, while white indicates the dot that is idle and occupied by an electron. The charging energy for each dot can be extracted from the diagrams. (c) and (d) Charge-stability diagrams for the array measured simultaneously by $\mathrm{SET_1}$ and $\mathrm{SET_2}$, respectively, where $\mathrm{vP1}$ is synchronized with $\mathrm{vP2}$ and $\mathrm{vP3}$ is synchronized with $\mathrm{vP4}$. The coefficients are calculated from the charging energies with $U_4$ as a reference. Before sweeping, the chemical potentials of dots 1 and 3 are intentionally set lower than those of dots 2 and 4, facilitating easier tunneling of electrons in dot 2(4) into or out. The staggered charge transitions in the diagrams are equivalent to overlapping (a) and (b) with an offset. By counting the charge transitions from the lower-left corner of the diagrams, the charge state in each regime can be determined, such as the single-electron occupancy state marked by the red stars. Each SET can detect charge transitions of all four dots but is more sensitive to their respective nearby dots. The disappearance of the transitions (white dashed lines) is due to the lower co-tunneling rates, particularly when the electrons in dots 1 or 3 are emptied.}
  \label{fig2:Figure}
\end{figure}

Having tuned the array to single-electron occupancy, a virtual gate space was established to compensate for the crosstalk and achieve independent control of the chemical potential of each dot. Further details about the virtual gate can be found in \supplementaryinfo~Section S3. \autoref{fig2:Figure}a-b show the charge-stability diagrams measured in virtual gate space for dots 1 and 2, and dots 3 and 4. Orthogonality in charge transitions indicates that independent control over each dot's chemical potential has been achieved. The absence of any charge transitions in the lower-left corner verified that electrons within each dot can be emptied. Moreover, these diagrams allow for the extraction of charging energy, a critical parameter for characterizing the reproducibility of quantum dots\cite{RN1341}, by measuring the spacing between charge transitions and converting it with a lever arm. The value is determined from the Coulomb diamonds measured in the finite voltage-bias transport experiments with a single quantum dot formed under each plunger gate. Here, we use the average lever arm $\alpha\, \mathrm{\approx 0.12 \, \mathrm{eV/V}}$ for simplicity, given that the difference among the four quantum dots is less than $0.01\, \text{eV/V}$. The charging energies for the four dots are $U_1=2.83 \text{ meV},\, U_2=5.02 \text{ meV},\, U_3=3.05 \text{ meV},\, \text{ and } U_4=4.63 \text{ meV}$. We observe that the charging energies for dots 1 and 3 are almost identical, as are those for dots 2 and 4, owing to the central symmetry of the gate layout. However, the anti-diagonal dots exhibit notably larger charging energies than the diagonal dots. This discrepancy can be explained by the relatively stronger electric potential confinement experienced by dots 2 and 4, resulting from the presence of corner-shaped screening gates (see \supplementaryinfo~Section S4).

The inhomogeneity of charging energies suggests the presence of significant electrostatic disorder, and addressing this obstacle is crucial for achieving homogeneous electron filling across the array\cite{RN1437}. Here, we involve a method to synchronize the sweeping of the four virtual plunger gates, guided by a coefficient determined from the charging energies. The scanning parameters used here are derived from the combination of plunger gates, such as $\mathrm{vP1}$ synced with $\mathrm{vP2}$ and $\mathrm{vP3}$ synced with $\mathrm{vP4}$. Before gate sweeping, the chemical potentials of dots 1 and 3 are intentionally set to be slightly lower than those of dots 2 and 4, aiming to demonstrate the electron filling of each dot more visually. As a result, a staggered pattern is observed in the charge-stability diagrams shown in \autoref{fig2:Figure}c-d, obtained by simultaneous readout of $\mathrm{SET_1}$ and $\mathrm{SET_2}$ (see \supplementaryinfo~Section S5 for diagrams under a wider voltage range). The nearly equally spaced charge transitions for each dot indicate that homogeneous electron filling in the array can be achieved. Moreover, the diagrams effectively encompass the charge transitions corresponding to all four dots, as well as the intersections between these transitions. By comparing the signals from both SETs, we observe that the interdot charge transitions for dots 1 and 4 are more pronounced in \autoref{fig2:Figure}c, as highlighted by dotted white circles, while those for dots 2 and 3 are more distinctly observed in \autoref{fig2:Figure}d, marked by dotted black circles. We utilize this distinction to differentiate quantum dots and subsequently determine the charge occupancy within each regime. Overall, this synchronized sweeping method allows for comprehensive visualization and analysis of both individual and collective dot behaviors solely from a 2D charge-stability diagram.

\begin{figure}[H]
  \centering
  \includegraphics{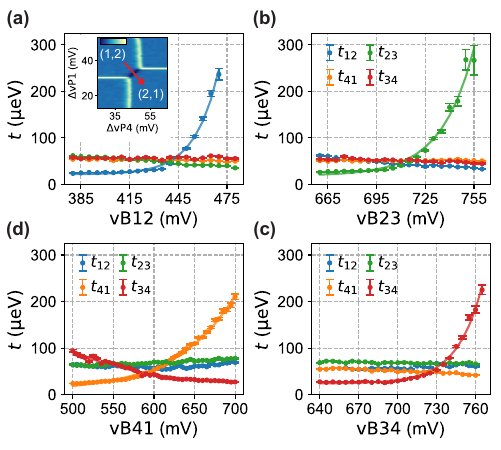}
  \caption{Controllable nearest-neighbor tunnel couplings. The inset in (a) shows an example of charge-stability diagram for dot pair 1-4, zoomed in near the $(1,2) \text{-} (2,1)$ interdot charge transition, and the adjacent dots are kept in the single electron Coulomb blockade regime. The red arrow indicates the detuning axis scan direction, perpendicular to the interdot transition. The extracted tunnel couplings are categorized and plotted as a function of the voltage applied on each virtual barrier gate $\mathrm{vB}{ij}$, as shown in (a)-(d). An exponential function is used to fit the data, as shown by the solid lines. Notably, when measuring the target tunnel coupling, all other unrelated dot pairs are configured in the weak coupling regime.}
  \label{fig3:Figure}
\end{figure}

Having achieved orthogonality control of the chemical potential of the array, we next demonstrate control over the interdot tunnel couplings. We first characterize the dependence between the tunnel couplings of nearest-neighbor dots and the corresponding barrier gates. To sequentially extract $t_{ij}$ at different barrier voltages, virtual barrier gate $\mathrm{vB}{ij}$ is used to avoid affecting the dot potentials. The results are shown in \autoref{fig3:Figure}a-d, plotted as a function of $\mathrm{vB}{ij}$, respectively. All interdot tunnel couplings, extracted by fitting polarization lines using the DiCarlo method \cite{RN10532,RN7400}, can be tuned over a wide range, from approximately $25 \ \mathrm{\mu eV}$ to beyond $200 \ \mathrm{\mu eV}$. The lower limit value can be attributed to thermal broadening ($150\ \mathrm{mK} \approx 13\ \mathrm{\mu eV}$), as well as back action from SETs and power broadening of AC excitations (see \supplementaryinfo~Section S6). Despite this, we can reasonably infer that the residual coupling is very small. The maximum measurable value is limited by the measurement accuracy, as increased tunnel coupling leads to significant charge hybridization, which blurs the interdot charge transitions.\cite{RN4305,RN4361}. 

As expected, $t_{ij}$ exhibits a strong exponential dependence on the corresponding $\mathrm{vB}{ij}$\cite{RN7347}. To quantitatively determine the tunability of barrier gate on tunnel coupling, an exponential function $t_{ij}=t_0+t_1\exp(\beta_{ij}\mathrm{vB}{ij})$ was used to fit the data, where $t_0$, $t_1$ and $\beta_{ij}$ are fitting parameters. The coefficients for all four $t_{ij}$ are $\beta_{12}=6.85\pm0.33\times10^{-2}\,\text{mV}^{-1} $, $\beta_{23}=4.96\pm0.45\times10^{-2}\,\text{mV}^{-1} $, $\beta_{34}=5.12\pm0.20\times10^{-2}\,\text{mV}^{-1} $, and $ \beta_{41}=1.42\pm0.05\times10^{-2}\,\text{mV}^{-1} $, respectively. Note that the tunability of $\mathrm{vB}41$ on $t_{41}$ is approximately 21\% of that of $\mathrm{vB}12$ on $t_{12}$, which implies that a wider voltage range is required to achieve the expected tunnel coupling strength, as evident from the comparison between \autoref{fig3:Figure}a and \autoref{fig3:Figure}d. However, we observed that the wide changes in voltage not only rendered the virtual gate ineffective but also induced significant displacement in the dot position, which, in turn, led to crosstalk between interdot tunnel couplings, as discussed below.

To extract the crosstalk between tunnel couplings of nearest-neighbor dots, we use a constant value of tunnel coupling, $t_{\mathrm{ref}}$, as a reference. Here the target coupling is configured at approximately $60 \, \mathrm{\mu eV}$, as clearly shown in \autoref{fig3:Figure}. Crosstalk can be quantified by measuring the change in $t_{ij}$ as the neighbor barrier gate voltage $\mathrm{vB}mn$ is varied. To facilitate comparison, the results are categorized by the $\mathrm{vB}ij$, as shown in \autoref{fig3:Figure}a-d. Over a wide voltage range up to 200 mV, all $t_{ij}$ values remain close to $t_{\mathrm{ref}}$, which is consistent with the observation in the 2D silicon dot array from ref. \citenum{RN11799}. However, there is an exception in our device, that is, $t_{34}$ decreases as $\mathrm{vB}41$ increases, which can be primarily explained by the unintended displacement of quantum dots induced by $\mathrm{vB}41$. This crosstalk can be compensated by adjusting $\mathrm{vB}34$ based on the extracted negative coefficient ($-1.03\pm0.11\times10^{-2}\,\text{mV}^{-1} $) to maintain $t_{34}$ unchanged, thereby achieving orthogonal control of tunnel couplings between all nearest-neighbor dots.

\begin{figure}[H]
  \centering
  \includegraphics{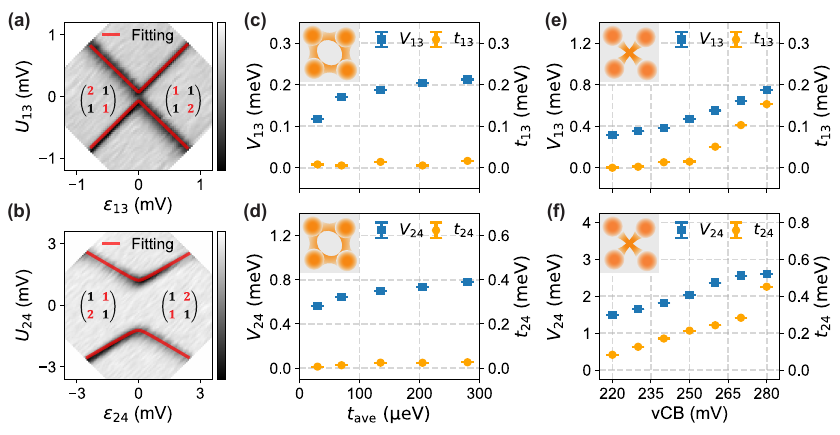}
  \caption{Controllable next-nearest-neighbor tunnel couplings. (a) and (b) Zoom in on the anti-crossings \colortext{in a virtual gate space of detuning and energy} for the diagonal and anti-diagonal dots, respectively. \colortext{Here detuning is defined as $\epsilon_{ij} = \alpha(\mathrm{vP}i - \mathrm{vP}j)$, and energy is defined as $U_{ij} = \alpha(\mathrm{vP}i + \mathrm{vP}j)$.} The sensor signal data is filtered using the threshold method (where 90\% of the data in the diagram is filtered), and then a hyperbolic function is employed to fit the filtered data with weights to obtain the inter-site interaction $V_{ij}$ and interdot tunnel coupling $t_{ij}$ (red lines). (c) and (d) $V_{ij}$ and $t_{ij}$ as a function of $t_\mathrm{ave}$ for diagonal and anti-diagonal dots, respectively, where $t_\mathrm{ave}$ indicates that all four $t_{ij}$ have nearly the same coupling strength. Control over $t_\mathrm{ave}$ is achieved by simultaneously increasing the voltage of virtual barrier gates corresponding to $t_{ij}$. The inset in (c) shows a schematic diagram of the coupling configuration of the array. (e) and (f) $V_{ij}$ and $t_{ij}$ as a function of CB gate voltage for diagonal and anti-diagonal dots, respectively. Here the virtual gate is used to change the CB gate voltage, and all nearest-neighbor dot pairs are maintained within the weak tunnel coupling regime, as shown by the inset in (e).}
  \label{fig4:Figure}
\end{figure}

In addition to nearest-neighbor tunnel couplings, we can also control the tunnel couplings between (anti-)diagonal dots. Neglecting the interactions with idle dots, the (anti-)diagonal dots here can be viewed as a weakly tunnel-coupled DQD described by a two-level system\cite{RN1816}. We extract the tunnel couplings by analyzing the anti-crossing features\cite{RN1437,RN11907} instead of interdot charge transitions, which are invisible for anti-diagonal dots due to their symmetrical location relative to SETs\cite{RN11908}. The shape and splitting of anti-crossings are determined by the tunnel coupling $t_{ij}$ and inter-site interaction $V_{ij}$ respectively, which can be obtained from hyperbolic fitting (the red solid lines in \autoref{fig4:Figure}a-b). See \supplementaryinfo~Section S7 for more details.

We first demonstrated that the tunnel couplings for both (anti-)diagonal dots can be completely turned off, which is verified by monitoring their changes while simultaneously increasing the couplings, $t_\mathrm{ave}$, for all four nearest-neighbor dots. \autoref{fig4:Figure}c-d show the extracted parameters for the diagonal and anti-diagonal dots, respectively, versus $t_\mathrm{ave}$. As $t_\mathrm{ave}$ increases from the weak coupling regime to approximately $300 \, \mathrm{\mu eV}$, both $t_{13}$ and $t_{24}$ remain unchanged, staying close to zero. This is further supported by the curvature evolution around the triple points (\supplementaryinfo~Section S8). From these consistent behaviors, we infer that the tunnel couplings of both diagonal and anti-diagonal dots are exceptionally weak, leaving only a residual coupling, which can be accurately determined by further experiments, such as exchange oscillations\cite{RN11714}. These findings are advantageous for silicon-based quantum computing, particularly for fault-tolerant coding schemes based on the surface code\cite{RN10693}. Moreover, the slight increase in $V_{24}$ and $V_{13}$ can be attributed to the tendency of quantum dots to move closer together as the barrier gate voltage increases. Notably, $V_{24}$ is nearly four times larger than $V_{13}$, primarily due to a stronger shielding effect from the screening gates, which pushes dots 2 and 4 closer together compared to dots 1 and 3 (\supplementaryinfo~Section S4).

To turn on the tunnel couplings of (anti-)diagonal dots, we need to increase the voltage of the CB gate. However, introducing the CB gate leads to a pronounced crosstalk effect, affecting not only the chemical potentials of each dot but also the tunnel couplings between them. The former can be effectively compensated for by the virtual gate, while for the latter, manual compensation is necessary and achievable by measuring the charge-stability diagrams of all pairwise dots. The method described in \autoref{fig3:Figure}a, in principle, can compensate for crosstalk on tunnel couplings, yet its effectiveness is greatly compromised due to the substantial impact of the CB gate. Moreover, we adhere to a pivotal guideline during tuning: ensure that the tunnel couplings of all nearest-neighbor dots remain in the weak coupling regime before proceeding to change the CB gate voltage. This strategy is beneficial for mitigating the crosstalk effect of the tunnel coupling of nearest-neighbor dots on the tunnel coupling of (anti-)diagonal dots.

The parameters extracted from the anti-crossings are shown in \autoref{fig4:Figure}e-f for the diagonal and anti-diagonal dots, respectively. A significant enhancement in the tunnel coupling is noticeable only when the CB gate voltage exceeds a certain threshold. Both $t_{13}$ and $t_{24}$ can be tuned from completely off to over $100 \, \mathrm{\mu eV}$, comparable to the nearest-neighbor tunnel couplings observed in \autoref{fig3:Figure}. Note that the tuning effect differs between the two pairs of dots, primarily due to the closer proximity of dots 2 and 4 compared to dots 1 and 3 (see \supplementaryinfo~Section S4). Therefore, by exploiting this discrepancy, we can keep $t_{13}$ in a closed state while activating $t_{24}$, thus controllably introducing a next-nearest-neighbor tunnel coupling term into the system. The charge-stability diagrams, in this case, are shown in \supplementaryinfo~Section S10 where $\mathrm{vCB=250\ mV}$. It should be noted that we can open $t_{24}$ while keeping $t_{13}$ closed, but cannot do the opposite, which is a limitation of our gate architecture.

Thus far, we have demonstrated high-level control over the electron filling of each dot, nearest-neighbor tunnel couplings, and particularly, next-nearest-neighbor tunnel couplings in our $2\times2$ QD array. This affords us the capability to tune quantum dot parameters entirely in an electrical manner, without the need for custom design and re-fabrication of devices\cite{RN11524,RN10935}. \added{More importantly, the selective control over tunnel couplings between nearest-neighbor and next-nearest-neighbor quantum dots is crucial for enabling arbitrary quantum simulations and computations.} We next illustrate this flexibility by demonstrating the ability to change the coupling configurations of the array. 

\begin{figure}[H]
  \begin{center}
    \includegraphics{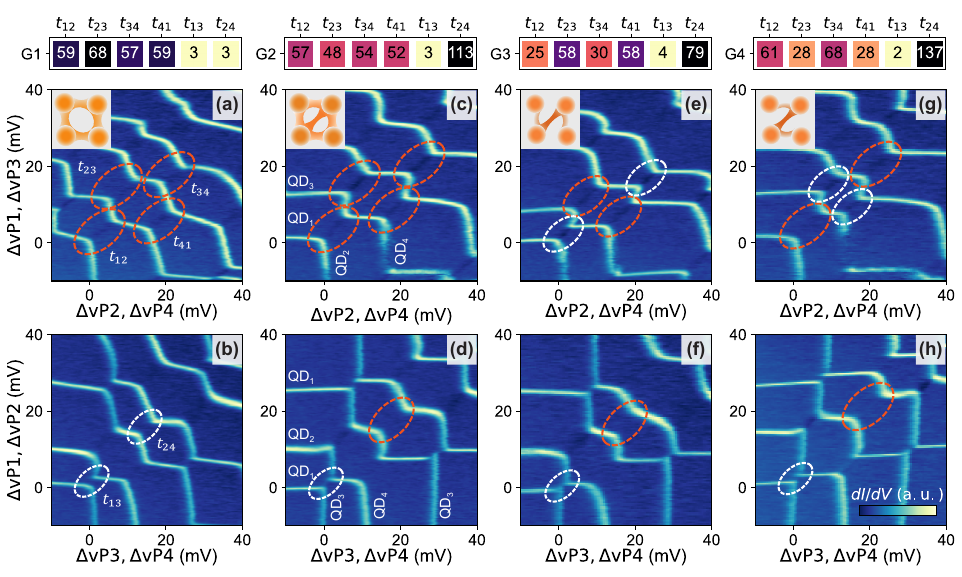} 
    \caption{Demonstration of different tunnel coupling configurations for the array. Different synchronized gates are swept to make certain anti-crossings visible in the diagrams. The first row shows the changes in nearest-neighbor tunnel couplings under different configurations, while the second row corresponds to next-nearest-neighbor tunnel couplings. Here the orange dashed circle indicates that the corresponding tunnel coupling is on, while the white dashed circle indicates that it is off. For each configuration, the tunnel coupling strengths ($\mathrm{\mu eV}$) for all dot pairs are shown at the top of the corresponding diagrams. \colortext{See \supplementaryinfo~Section S11 for more fiting details.} Here, the signals from two charge sensors are combined to enhance the contrast, where $ dI/dV = dI_{S1}/dV_{P2} + dI_{S1}/dV_{P4} + dI_{S2}/dV_{P2} + dI_{S2}/dV_{P4}$. (a) and (b) Charge-stability diagrams of a square lattice with all nearest-neighbor couplings on, while $t_{13}$ and $t_{24}$ approaching zero. (c) and (d) Charge-stability diagrams of an isotropic triangular lattice with all nearest-neighbor couplings and $t_{24}$ on. (e)-(h) Charge-stability diagrams of two sawtooth lattices with asymmetric couplings, where the couplings along the vertical or horizontal direction are off.}
    \label{fig5:Figure}
  \end{center}
\end{figure}

To visualize the array's filling and coupling states, we utilized the synchronized sweeping method described in \autoref{fig2:Figure}. \autoref{fig5:Figure}a shows the charge-stability diagram with CB gate voltage at zero, obtained by sweeping the synchronized gate combinations $\mathrm{vP1}$, $\mathrm{vP3}$ versus $\mathrm{vP2}$, $\mathrm{vP4}$. In this diagram, the anti-crossings are visible for all nearest-neighbor dot pairs (orange circles), with the coupling strengths shown at the top of \autoref{fig5:Figure}a. Similarly, by switching the synchronized gate combinations to $\mathrm{vP1}$, $\mathrm{vP2}$ versus $\mathrm{vP3}$, $\mathrm{vP4}$, we can observe the anti-crossings of (anti-)diagonal dots (white circles), as shown in \autoref{fig5:Figure}b. Here, $t_{13}$ and $t_{24}$ are extracted to be close to zero. This configuration holds great appeal for fault-tolerant quantum computing based on surface code requiring only nearest-neighbor couplings\cite{RN10693}. Additionally, it facilitates quantum simulation experiments, such as Nagaoka ferromagnetism\cite{RN7349} and resonating valence bond states\cite{RN11717}.

We then increase the CB gate voltage to 250 mV to turn on $ t_{24}$, thereby transforming the QD array from a square lattice to an isotropic triangular lattice, as shown in \autoref{fig5:Figure}c-d. The coupling strength between nearest-neighbor dots remains at the same level as in \autoref{fig5:Figure}a. The $t_{24}$ is extracted to be approximately 113 $\mathrm{\mu eV}$, while $t_{13}$ remains close to zero. These results demonstrate the ability to controllably introduce next-nearest-neighbor coupling into QD arrays, facilitating the exploration of frustration physics, including geometric frustration\cite{RN11602} and frustration-induced magnetism\cite{RN11845}. Furthermore, we can selectively deactivate the nearest-neighbor couplings to transform the system into sawtooth lattices with asymmetric interactions. The results are shown in \autoref{fig5:Figure}e-f, with $t_{12}$ and $t_{34}$ turned off, and in \autoref{fig5:Figure}g-h, with $t_{23}$ and $t_{41}$ turned off. These coupling configurations provide the possibility to mimic nontrivial phase physics in condensed matter systems upon scaling to larger arrays, such as the many-body Su-Schrieffer-Heeger (SSH) model\cite{RN10935,RN11787} and the zigzag ladder model\cite{RN11977,RN11923}. \added{Note that the capability for selective control over the next-nearest-neighbor couplings has certain limitations; specifically, it currently enables control over only one diagonal coupling.}

In conclusion, we have demonstrated a 2D quantum dot array in planar silicon that can be operated at single-electron occupancy and enables \deleted{selective} control over a wide range of tunnel couplings, including both nearest-neighbor and next-nearest-neighbor couplings. We employed a synchronized sweeping method to obtain a comprehensive representation of all quantum dots from a single 2D plot, illustrating the orthogonal and homogeneous electron filling of the array. Leveraging the remarkable tunability, the coupling configurations of the array can be changed from a square lattice to an isotropic triangular lattice, and even sawtooth lattices. These results underscore the potential of silicon quantum dots as versatile platforms with high controllability and connectivity for advancing quantum computing and quantum simulation. Furthermore, future experiments, such as photon-assisted tunneling\cite{RN1398} and exchange oscillations\cite{RN11714}, will be crucial for determining the precise value of the residual coupling. To address the limitations imposed by the screening gates and facilitate scaling up to larger arrays, such as $2 \times N$ and $N \times N$, it may be beneficial to change the dot arrangement from a square to a rhombus-like configuration.\cite{RN12057} This arrangement could reduce the distance between anti-diagonal dots, potentially allowing for more effective control over their coupling while suppressing the coupling between diagonal dots. The challenges of low yield, poor uniformity, and planar gate fan-out in scaling are expected to be addressed by leveraging industrial semiconductor manufacturing techniques\cite{RN10467,RN10696,RN11981}.

\begin{suppinfo}
  \noindent
  The supporting information is available free of charge via the internet at \url{http://pubs.acs.org}.
  Device design and fabrication details; consecutive tuning procedure and versatility demonstration; capacitive crosstalk matrix; charge-stability diagrams under a wide voltage range; electric potential simulation; evolution of charge-stability diagrams under $t_\mathrm{ave}$; evolution of anti-crossings for (anti-)diagonal dots under $\mathrm{vCB}$; and charge-stability diagrams at $\mathrm{vCB=250\, mV}$.
\end{suppinfo}

\begin{acknowledgement}
  \noindent
  This work was supported by the Innovation Program for Quantum Science and Technology (Grant No. 2021ZD0302300) and the National Natural Science Foundation of China (Grants No. 92265113, No. 12074368, No. 12034018 and No. 92165207). This work was partially carried out at the University of Science and Technology of China Center for Micro and Nanoscale Research and Fabrication.
\end{acknowledgement}

\bibliography{achemso-demo}
\end{document}